\documentclass[aps,pre,twocolumn]{revtex4}
\usepackage{amssymb}

\usepackage{eurosym}
\usepackage{epsfig}

\begin{document}

\title{Folding of the  frozen-in-fluid di-vorticity field  in two-dimensional hydrodynamic turbulence}

\author{E.A. Kuznetsov$^{a,b,c}$\/\thanks{%
kuznetso@itp.ac.ru} and E.V. Sereshchenko$^{c,d,e}$}
\affiliation{
{\small \textit{$^{b}$ P.N. Lebedev Physical Institute, RAS, 119991 Moscow, Russia}}\\
{\small \textit{$^{c}$ L.D. Landau Institute for Theoretical Physics, RAS, 119334 Moscow, Russia}}\\
{\small \textit{$^{a}$ Novosibirsk State University, 630090 Novosibirsk, Russia}}\\
{\small \textit{\ $^{d}$A.S. Khristianovich Institute of Theoretical and Applied Mechanics, SB RAS, 630090 Novosibirsk, Russia}}\\
{\small \textit{$^{e}$ Far-Eastern Federal University, 690950 Vladivostok, Russia }}}

\begin{abstract}
The vorticity rotor field ${\bf B}=\mbox{rot}\,\mathbf{\omega}$ (di-vorticity) for freely decaying two-dimensional hydrodynamic turbulence due to a tendency to breaking is concentrated in the vicinity of the lines corresponding to the position of the vorticity quasi-shocks. The maximum value of the di-vorticity $B_{max}$ at the stage of quasi-shocks formation increases exponentially in time, while the thickness $\ell(t)$ of the maximum area in the transverse direction to the vector ${\bf B}$ decreases in time also exponentially. It is numerically shown that $B_{max} (t)$ depends on the thickness according to the power law: $B_{max}(t)\sim \ell^{-\alpha}(t)$, where the exponent  $\alpha\approx 2/3$. This behavior indicates in favor of  folding for the divergence-free vector field of the di-vorticity.
\end{abstract}

\maketitle

\vspace{0.2 cm}
PACS: {52.30.Cv, 47.65.+a, 52.35.Ra}

\section{Introduction}

In the two-dimensional developed hydrodynamic turbulence at  high Reynolds numbers, $Re \gg 1$, as it was found out in  \cite{KNNR-07, KKS, KS-15}, the Kraichnan direct cascade  \cite{kraichnan} with a constant enstrophy flux  is formed due to the appearance of the vorticity quasi - shocks (or jumps), because of the compressibility of continuously distributed lines of the field of the vorticity rotor,  ${\bf B}=\mbox{rot}\,\mathbf{\omega}$, 
often called the di-vorticity after Kida \cite{kida}. This compressibility property follows directly from the equation for ${\bf B}$,
\begin{equation} \label{Helmholtz}
\frac{\partial {\bf B}}{\partial t} =\mbox{rot}[{\bf v}\times {\bf B}],\,\,\mbox{div}\,{\bf v}=0,
\end{equation}
which has the form of a frozenness equation. From this equation  we have that ${\bf B}$ changes only  by virtue of the velocity component ${\bf v_n}$ normal to the di-vorticity vector (in consequence of the vector product). The velocity component ${\bf v_n}$ due to frozenness ${\bf B}$ determines the change in the position of the force lines of this field by means of the  Lagrangian trajectories, which are defined from the equations
\[
\frac{d{\bf r}}{d t} ={\bf v_n}({\bf r},t)=0,\,\,\, 
{\bf r}|_{t=0}={\bf a}.
\] 
In the general situation, $\mbox{div}\,{\bf v_n}\neq 0$, and therefore the mapping ${\bf r}={\bf r}({\bf a},t)$ as the solution of these equations is turned out to be compressible. This fact also follows from the Liouville equation for the mapping Jacobian  $J=\partial(x,y)/\partial(a_x,a_y)$ (as a measure of the  infinitesimally small area  variation):
\[
\frac{d{J}}{d t} =\mbox{div}\,{\bf v_n}\cdot J.
\]
Thus,  Jacobian $J$ can take arbitrary values, including zero. This is the reason for the compressibility of continuously distributed di-vorticity lines and, accordingly, the tendency to breaking, that results in the formation of vorticity quasi-shocks.

In the case of freely decaying turbulence, this process is dominant, leading to a strong anisotropy of the turbulence spectrum because of  the presence of jets generated by quasi-shocks \cite{KNNR-07, KKS}. This process turns out to be the fastest, as a result the turbulence spectrum of the direct cascade at the initial stage forms a power dependence on the wave number $k$ with the Kraichnan type behavior  for the spectrum: $E_k\sim k^{-3}$ (see the original paper by Kraichnan \cite{kraichnan}), even in the presence of pumping, as shown by numerical experiments \cite{KS-17}. At the same time, the formation of the vorticity quasi-shocks  is exponential; in accordance with this, the regions of the maximum of di-vorticity are decreased in the direction perpendicular to the lines of the constant vorticity. As demonstrated in the numerical experiments \cite{KNNR-07, KKS, KS-15}), for typical initial conditions an increase of the di-vorticity modulus consists 2-2.5 orders of magnitude, and the transverse size of the maximal area ${\bf B}$ decreases significantly. The explanation of this growth is related to the possibility of partial integration of the equation (\ref{Helmholtz}) in terms of mapping ${\bf r}={\bf r} ({\bf a},t)$ \cite{KNNR-07}:
\begin{equation} \label{B}
\mathbf{B}(\mathbf{r},t) = \frac{(\mathbf{B}_0(\mathbf{a})
	\cdot\nabla_a)\mathbf{r}(\mathbf{a},t)}{J},
\end{equation}
where $\mathbf{B_0} (\mathbf{a})$ is
the initial  $\mathbf{B}$, which has a meaning  of the Cauchy invariant analogue. A similar formula for the three-dimensional Euler equations is basic in the so-called vortex line representation  \cite{KR}. The key point here for understanding is the compressibility of the di-vorticity field and the possibility of $J$ to vanish. As is known, breaking in the gas dynamics occurs due to the compressibility of the gas. The formation of quasi-two-dimensional caustics occurs when approaching the breaking point (see, e.g. \cite{zeldovich}). The formation of the vorticity quasi-shocks happens similarly. It is necessary to mention that probably first time appearance of the vorticity  quasi-shocks was observed in the numerical simulations by Lilly in 1971 \cite{Lilly}. It was one of the main motivation for Saffman to suggest the spectrum $E(k)\sim k^{-4}$ \cite{saffman} different from that suggested by  Kraichnan \cite{kraichnan}. The first explanation of these facts was given in the papers \cite{KNNR-07} based on the representation (\ref{B}) in 2007. 

In this paper, we investigate how the maximum value of the di-vorticity varies depending on the thickness of the maximum area in order to find out whether this process can be considered as a fold formation (what is a fold - see, for example, \cite{Arnold}).  As a result of numerical simulation on the grid 16384x16384, we found that between the maximum value of $B_{max}$ and the thickness of $\ell$, at the stage of exponential growth, there is a power law dependence: $B_{max}\sim \ell^{\alpha}$ , where the exponent $\alpha$ is close to $2/3$. It should be noted that such a dependence was found while formation of the pancake-type vortex structures arising for inviscid three-dimensional flows \cite{AKM-18}. This result indicates that the formation of quasi-shocks can be considered as a process of folding for a divergent-free vector field - the di-vorticity field. If for the three-dimensional Euler equation, the appearance of a power dependence between the maximum vorticity $\omega_{max}$ and the pancake thickness $\ell$ of the form $\omega_{max}\sim \ell^{-2/3}$ could be attributed to the Kolmogorov type relation, then the dependence $B_{max}\sim \ell^{-2/3}$ indicates that in the two-dimensional Euler hydrodynamics we are dealing with folding that is not related to the Kolmogorov' behavior.The frozenness of both the vorticity field in the three-dimensional Euler equation and the di-vorticity field combines both of these cases - the formation of three-dimensional vortex structures of pancake type and quasi-shocks of vorticity for two-dimensional flows. Despite the incompressibility of the velocity field, both fields - three dimensional vorticity field and di-vorticity are compressible (see \cite{AKM-18}). Apparently, due to this property of frozen-in-fluid fields, we can expect that the $2/3$ law is universal for any fields of this type.

\section{Numerical results}
In this paper for the numerical integration of the two-dimensional Euler equations written in terms of vorticity (Helmholtz equation),
\begin{equation} \label{euler}
\frac{\partial \omega}{\partial t} +({\bf v}\nabla) \omega=0,
\end{equation}
\begin{figure}[b!]
	\label{fig1}
	\centerline{
		\includegraphics[width=0.45\textwidth]{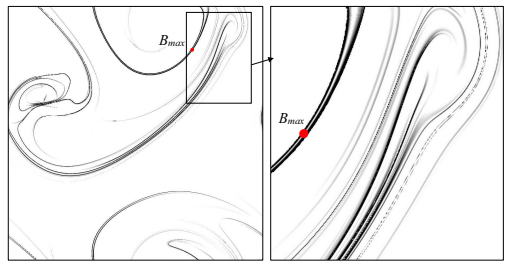}}
	\caption{Distribution of di-vorticity $|B|$ at $t=12$.}
\end{figure}
\begin{figure*}[t!]
	\centerline{ 
		\includegraphics[width=0.45\textwidth]{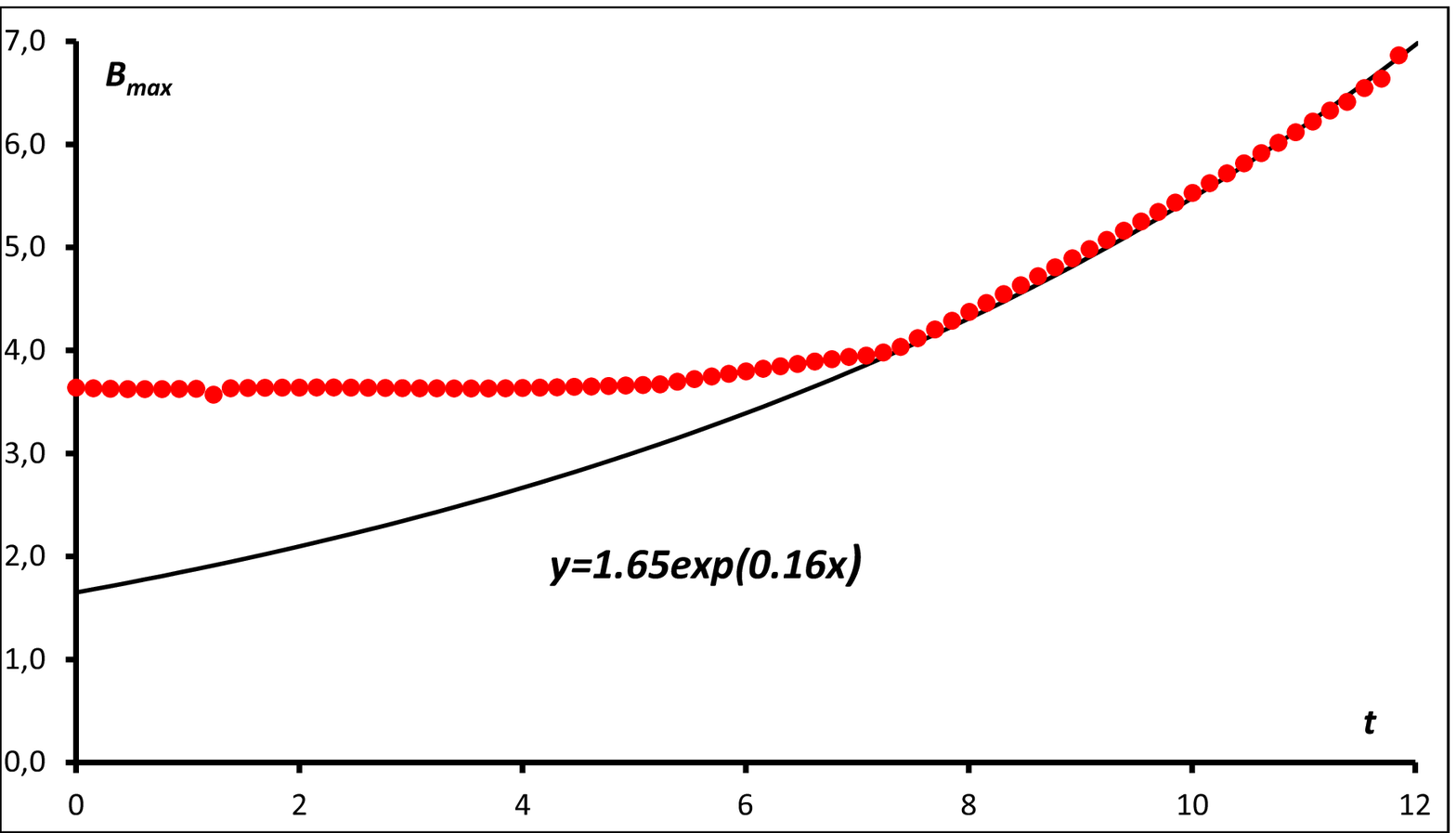}
		\includegraphics[width=0.45\textwidth]{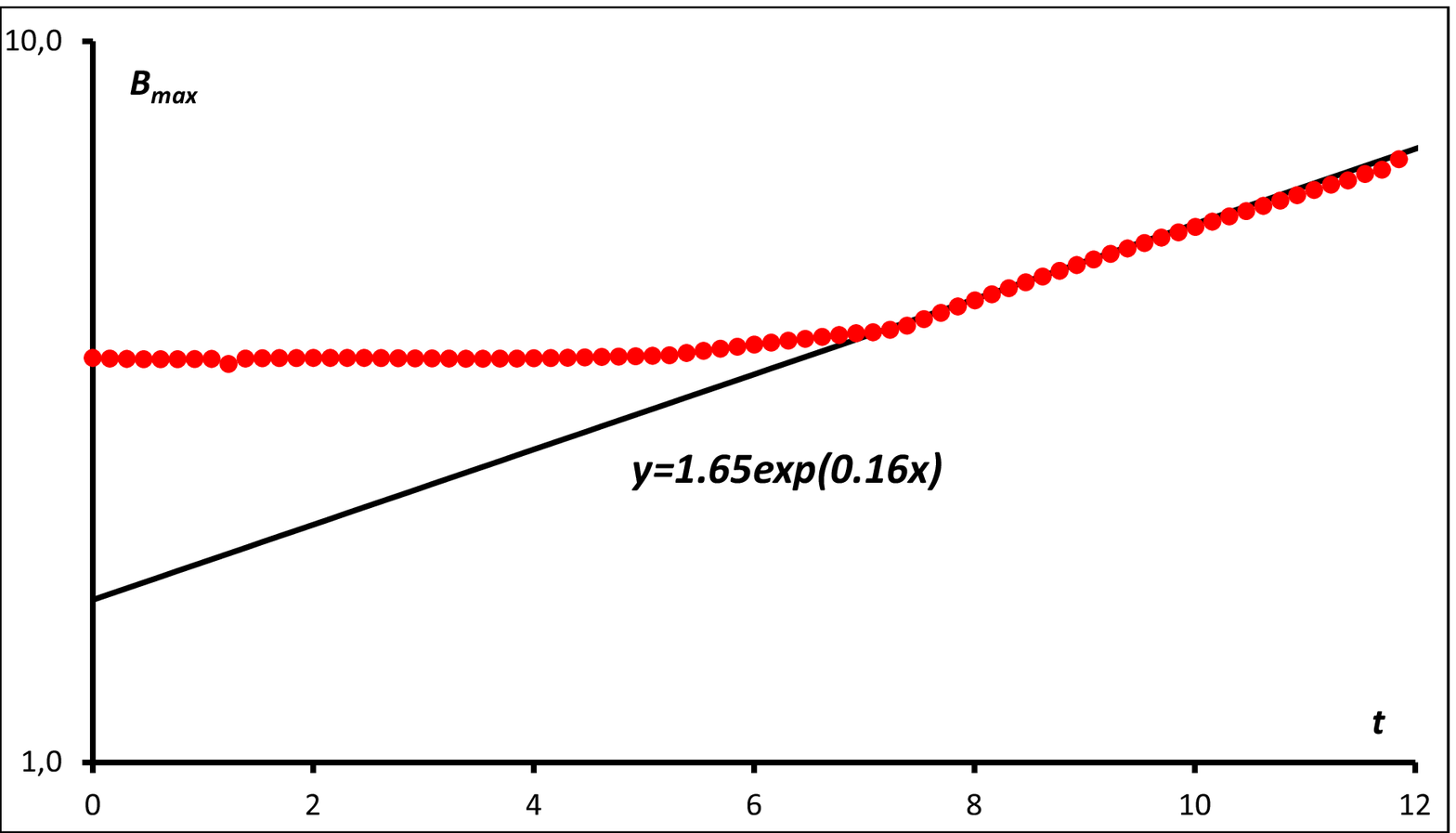}
	}
	\caption{The dependence of the maximum di-vorticity on time (on the left is the usual scale, on the right is the logarithmic). The points correspond to the numerical results, and the line to exponential fitting.}
	\label{fig2}
\end{figure*}
\begin{figure*}[]
	\centerline{ 
		\includegraphics[width=0.45\textwidth]{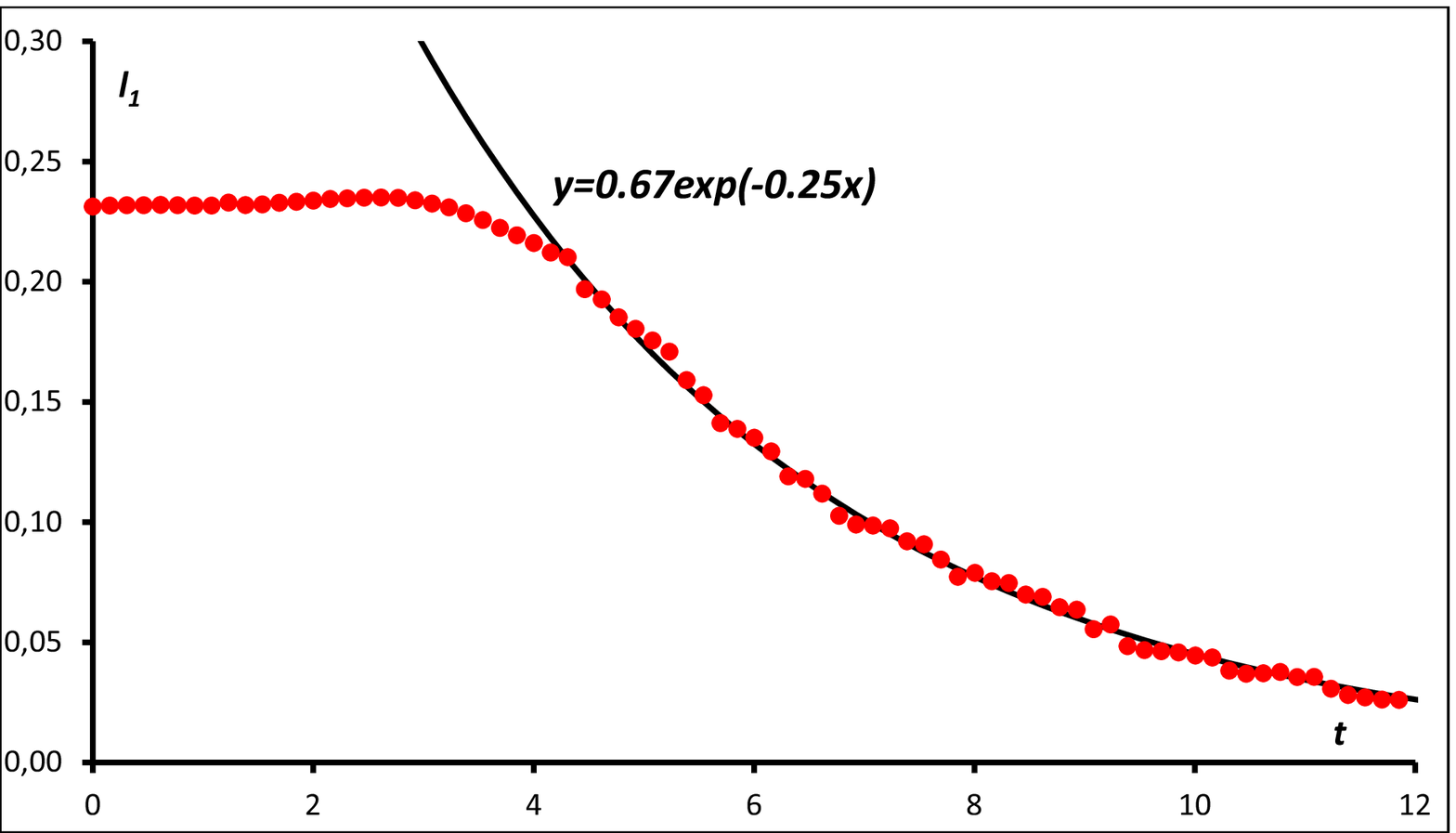}
		\includegraphics[width=0.45\textwidth]{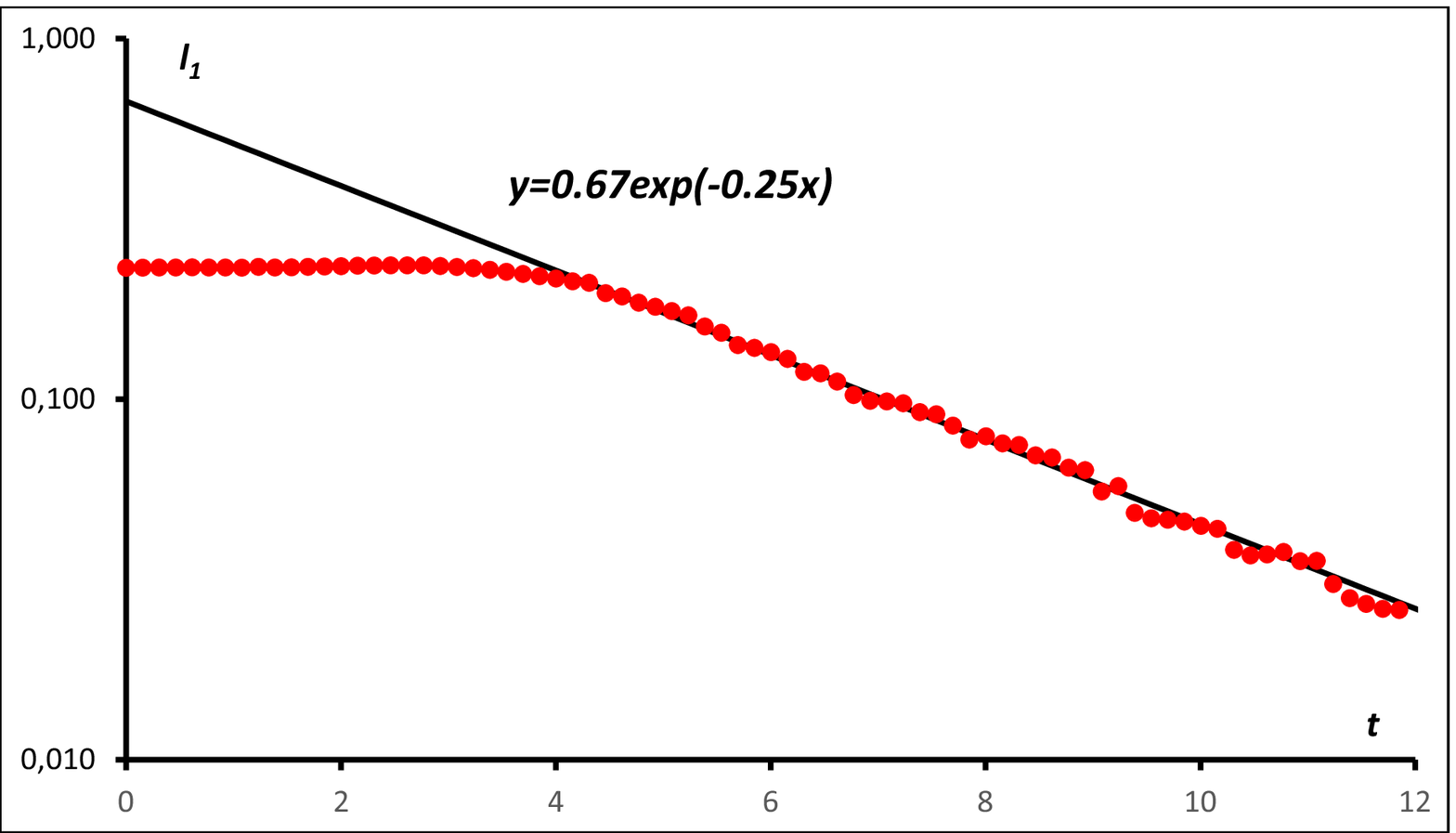}
	}
	\caption{Dependence of thickness $\ell_1$ on time (on the left is the usual scale, on the right is the logarithmic). The points correspond to the numerical results, and the line to exponential fitting.}
	\label{fig3}
\end{figure*}
\begin{figure*}[]
	\centerline{ 
		\includegraphics[width=0.45\textwidth]{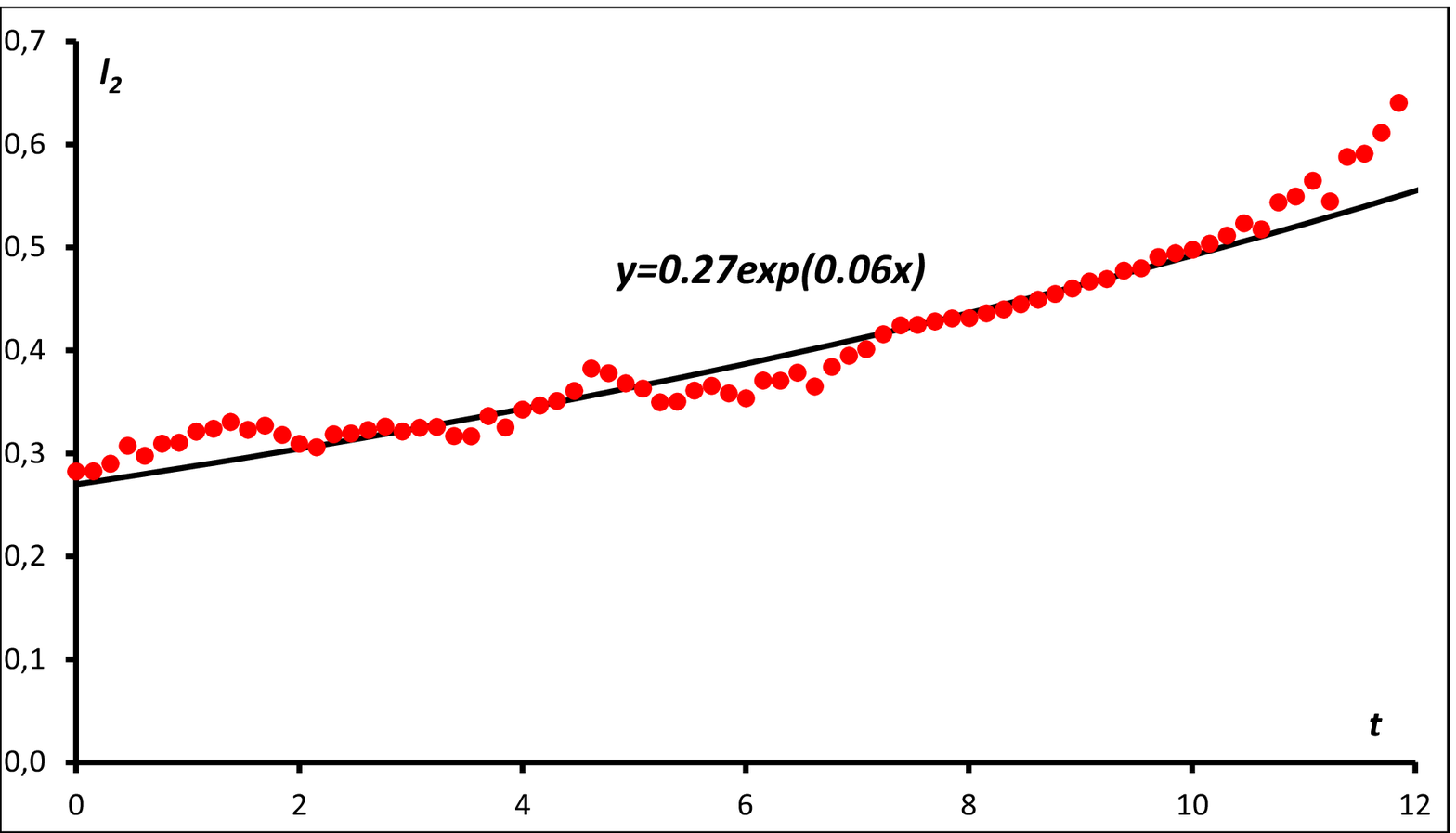}
		\includegraphics[width=0.45\textwidth]{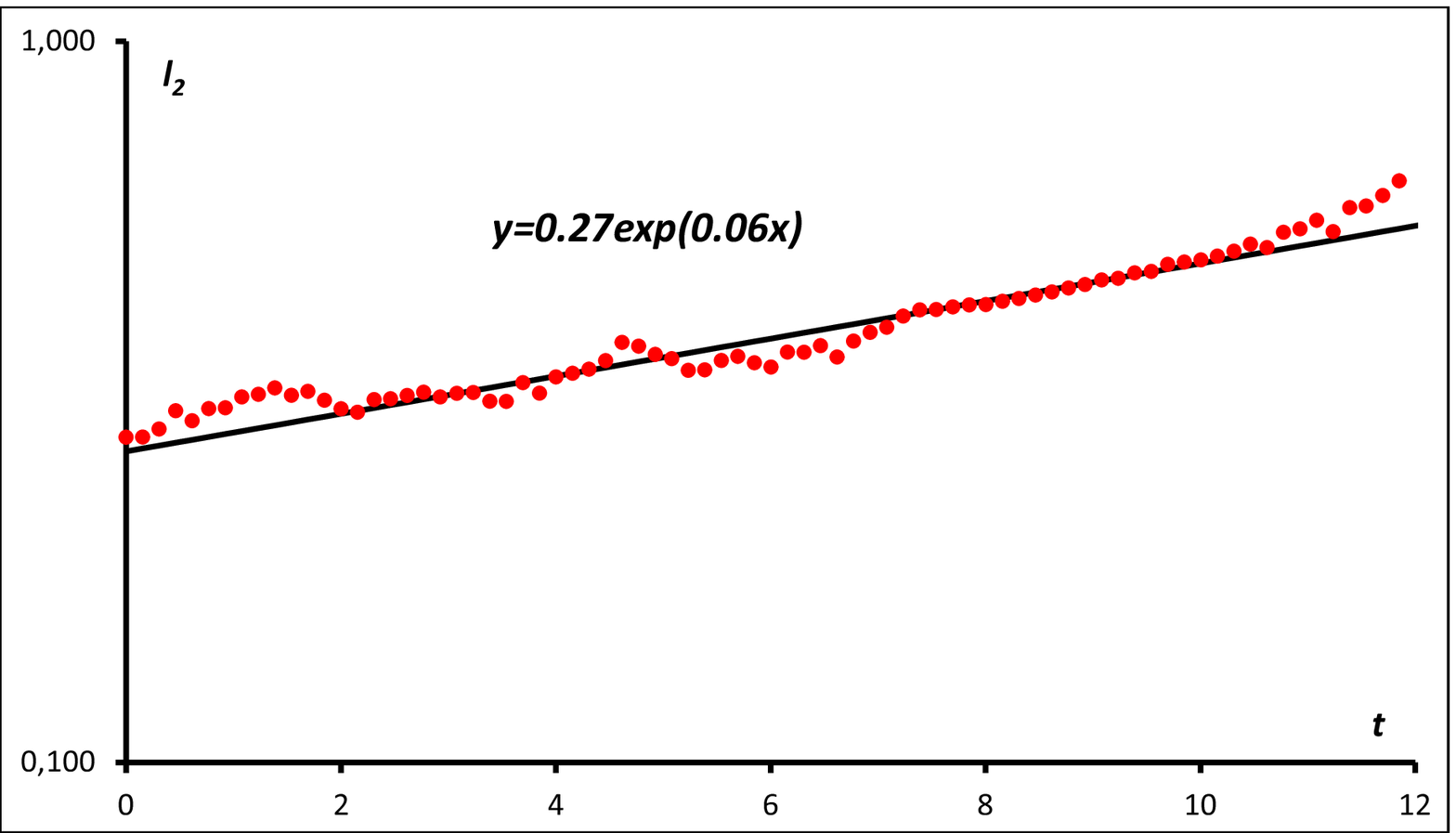}
	}
	\caption{Dependence of the longitudinal size $\ell_2$ on time (on the left is the usual scale, on the right is the logarithmic). The points correspond to the numerical results, and the line to exponential fitting.}
	\label{fig4}
\end{figure*}
\begin{figure}[b]
	\label{fig5}
	\centerline{
		\includegraphics[width=0.45\textwidth]{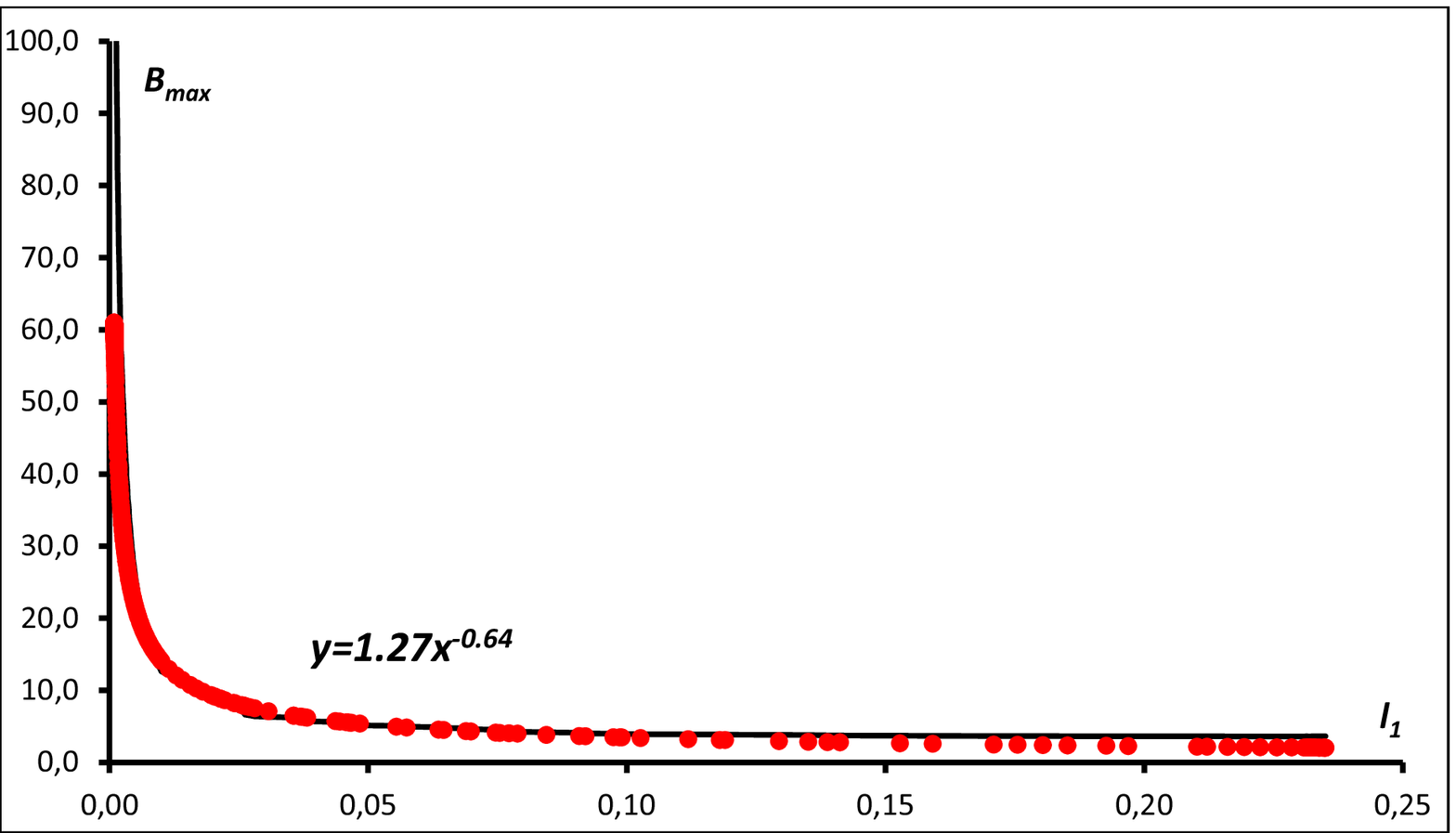}}
	\caption{The dependence of the maximum di-vorticity on the thickness $\ell$. The points correspond to numerical results, and the line corresponds to the power dependence $B_{max}\sim \ell^{-2/3}$.}
\end{figure}
we use algorithm described in detail in our previous papers \cite{KKS, KS-15, KS-17}. The integration domain for the equation (\ref{euler}) was a square box with sizes $2\pi \times 2\pi$, the boundary conditions were periodic at both coordinates. Velocity ${\bf v}$ and vorticity $\omega$ were found through the stream function $\psi$ using the standard formulas. For a given value of $\omega$ the stream function was found by reversing the Laplace operator with the help of the fast Fourier transform and then the velocity was determined. As in the our  previous works, two sets of Gaussian vortices with positive and negative vorticity with zero total vorticity were used as initial conditions. The size of each pair was random in the range of 0.2 - 0.6, the location of vortices was also random.  Unlike previous works  \cite{KKS, KS-15}, we limited the number of vortices to 8 (4 positive and 4 negative) to more accurately determine the required dependences - the ${\bf B}$ field and its geometric characteristics: maximum positions, longitudinal and transverse quasi-shock sizes, etc.

At first, the spatiotemporal dependences of vorticity were numerically found and then the temporal evolution of the field of di-vorticity was determined. Analysis of the distribution of the di-vorticity field showed that its main mass is concentrated in the small vicinity of the lines with a maximum value of $|B|$ in the form of narrowing two-dimensional ribbons that form a complex net and at large times becomes turbulent. Fig. 1 shows the structure of $|B|$ at $t=12$. Zoom shows that between the maximum lines of di-vorticity values of $|B|$ is significantly less than the maximum. For vorticity this corresponds to a terrace with steps of variable height. Each of these steps is a vorticity quasi-shock. As already noted, the maximum amplitude of the di-vorticity $B_{max}$ at the initial stage increases exponentially and then after reaching its maximum performs small oscillations near this maximum. Fig. 2 shows the dependence of $B_{max}$ on time at the initial stage. As can be seen from these figure, exponential growth is observed at times from $t=4$ to $t=12$ with the growth rate $0.16$.
To find the transverse size in the vicinity of the maximum $B_{max}$, the Hessian matrix $\partial_i\partial_j|B|$ was calculated at the maximum point and then by its eigenvalues the transverse (relative to the ribbon) size $\ell_1\, (\equiv \ell) $ and longitudinal $\ell_2$ were defined as $\ell_i=(2|\lambda_i|/B_{max})^{1/2}$, where $\lambda_i$ are the eigenvalues of the $\partial_i\partial_j|B|$ matrix. Fig. 3 shows the dependence of the thickness $\ell_1$ on time: at first $\ell_1$ practically does not change and then starting with $t=4$ there is an exponential decrease to $t=14$ with the negative growth rate $-0.25$.  At $t=14$ the number of grid points on the thickness of maximal di-vorticity line was about $20$. For longer times, this number decreases. So at $t=20$ the thickness reaches a value of $0.00045$, which is comparable to the grid size $2\pi / 16384=0.00038$, i.e. the further analysis becomes already incorrect. It should be noted that for $t>14$ exponential growth of $B_{max}$ stops and saturation takes place accompanying by small oscillations (see ref. \cite{KNNR-07, KKS}).

As for the longitudinal scale $\ell_2$, it grows exponentially up to $t=14$ with the growth rate $0.06$ (see Fig. 4). At large times $\ell_2$ growth stops, and then there is even an exponential fall, which in our opinion is associated with the discreteness of the computation grid and therefore is incorrect.

It is important to note that up to $t=14$ the total value of the growth rates (for both $\ell_1$ and $\ell_2$) turns out to be negative: $-0.25+0.06=-0.19$, which indicates the compressibility of the region of the maximal value $B_{max}$. Similar behavior is observed in three-dimensional hydrodynamics when forming pancake-type vortex structures \cite{AKM-15}. 

The obtained dependences for $B_{max}$ and thickness $\ell$ show that at the exponential stage between these values there is a power dependence $B_{max}=C \ell^{\alpha}$ with $\alpha= 0.16/(-0.25) = -0.64 \approx -2/3 $, $C$ is a constant (see Fig. 5).

It is worth noting  that this dependence of $B_{max}$ on $\ell$ in the form of the $2/3$ law was also verified  for another initial conditions (recall that the positions of vortices and their sizes were random). This allows ones to believe that this relation can be considered as universal.

\section{Conclusion}
The main conclusion of this work is that the formation of the power dependence of $B_{max}$ on the thickness $\ell$ - the  $2/3$ law - can be considered as folding, we emphasize, for the divergence-free vector field of the di-vorticity. As already noted in the introduction, the same scaling was found for the formation of the pancake-type vortex structures for three-dimensional Euler hydrodynamics \cite{AKM-18, AKM-15}. In this case, the scaling $2/3$ was set between the maximal vorticity and the pancake thickness: $\omega_{max}\sim\ell^{-2/3}$. Both these phenomena are united by the frozenness property of the vorticity field for the three-dimensional flows and  the di-vorticity for two-dimensional hydrodynamics. Is this law universal for any frozen-in-fluid fields for incompressible flows?
This question remains still open; especially it is interesting for magnetic hydrodynamics where, in the non-dissipative limit, the magnetic field  is also frozen.

This work was carried out with the support of the Russian Foundation of Basic Research (grant no. 17-01-00622).

\end{document}